\documentclass[3p]{elsarticle}
\UseRawInputEncoding
\usepackage{lineno,hyperref}
\usepackage{graphicx}
\usepackage{amsmath}
\usepackage{color}
\usepackage{mathtools}
\usepackage[normalem]{ulem}
\usepackage{rotating}
\usepackage{lscape}

\journal{arXiv}
\begin{document}

\begin{frontmatter}

\title{Are there Dragon Kings in the Stock Market?}

\author[mymainaddress]{Jiong Liu}
\author[mymainaddress]{M. Dashti Moghaddam}
\author[mymainaddress]{R. A. Serota\fnref{myfootnote}}
\fntext[myfootnote]{serota@ucmail.uc.edu}

\address[mymainaddress]{Department of Physics, University of Cincinnati, Cincinnati, Ohio 45221-0011}

\begin{abstract}
We undertake a systematic study of historic market volatility spanning roughly five preceding decades. We focus specifically on the time series of realized volatility (RV) of the S\&P500 index and its distribution function. As expected, the largest values of RV coincide with the largest economic upheavals of the period: Savings and Loan Crisis, Tech Bubble, Financial Crisis and Covid Pandemic. We address the question of whether these values belong to one of the three categories: Black Swans (BS), that is they lie on scale-free, power-law tails of the distribution; Dragon Kings (DK), defined as statistically significant upward deviations from BS; or Negative Dragons Kings (nDK), defined as statistically significant downward deviations from BS. In analyzing the tails of the distribution with $RV > 40$, we observe the appearance of "potential" DK which eventually terminate in an abrupt plunge to nDK. This phenomenon becomes more pronounced with the increase of the number of days over which the average RV is calculated -- here from daily, $n=1$, to "monthly," $n=21$. We fit the entire distribution with a modified Generalized Beta (mGB) distribution function, which terminates at a finite value of the variable but exhibits a long power-law stretch prior to that, as well as Generalized Beta Prime (GB2) distribution function, which has a power-law tail. We also fit the tails directly with a straight line on a log-log scale. In order to ascertain BS, DK or nDK behavior, all fits include their confidence intervals and p-values are evaluated for the data points to check if they can come from the respective distributions.
\end{abstract}


\end{frontmatter}

\section{Introduction\label{Introduction}}

Realized volatility $RV$ is the square root of realized variance, which is defined as follows

\begin{equation}
RV^2=100^2\times\frac{252}{n}\sum_{i=1}^nr_i^2
\label{RV2}
\end{equation}
where
\begin{equation}
\overline{r_n^2} = \frac{1}{n}\sum_{i=1}^nr_i^2
\label{RV2av}
\end{equation}
is the average realized variance over $n$ days and
\begin{equation}
r_i=\ln\frac{S_{i}}{S_{i-1}}
\label{ri}
\end{equation}
are the daily returns with $S_{i}$ being the reference (closing) price on day $i$. This is an annualized value, where $252$ represents the number of trading days in a year. In particular, $n=1$ represent daily returns and $n=21$, being a typical number of trading days in a month, is useful for evaluating monthly RV. We point out however that in our calculation $n$ is simply a number of consecutive trading days that can fall on different weeks and months. Specifically, we performed our analysis for $n=1, 2, 3, 5, 7, 9, 13, 17, 21$. Here we present results for $n=1, 7, 21$ which already succinctly illustrate the changes in the RV distribution with $n$.

\pagebreak
\newpage
Since it is based on actual trades, realized volatility (RV) is the ultimate measure of market volatility, although the latter is more often associated with the implied volatility, most commonly measured by the VIX index \cite{cboevix,cboevixhistoric} -- the so called market "fear index" -- that tries to predict RV of the S\&P500 index for the following month. Its model-independent evaluation \cite{demeterfi1999guide} is based on options contracts, which are meant to predict future stock prices fluctuations \cite{whitepaper2003cboe}. The question of how well VIX predicts future realized volatility has been of great interest to researchers \cite{christensen1998relation, vodenska2013understanding, kownatzki2016howgood, russon2017nonlinear}. Recent results \cite{dashti2019implied,dashti2021realized} show that VIX is only marginally better than past RV in predicting future RV. In particular, it underestimates future low volatility and, most importantly, future high volatility. In fact, while both RV and VIX exhibit scale-free power-law tails, the distribution of the ratio of RV to VIX also has a power-law tail with a relatively small power exponent \cite{dashti2019implied,dashti2021realized}, meaning that VIX is incapable of predicting large surges in volatility.

It should be emphasized that RV is agnostic with respect to gains or losses in stock returns. Nonetheless, it has been habitual that large gains and losses occur at around the same time. Here we wish to address the question of whether the largest values of RV fall on the power-law tail of the RV distribution. As is well known, the largest upheavals in the stock market happened on, and close to, the Black Monday, which was a precursor to the Savings and Loan crisis, the Tech Bubble, the Financial Crisis and the COVID Pandemic. Plotted on a log-log scale, power-law tails of a distribution show as a straight line. If the largest RV fall on the straight line they can be classified as Black Swans (BS). If, however, they show statistically significant deviations upward or downward from this straight line, they can be classified as Dragon Kings (DK) \cite{sornette2009,sornette2012dragon} or negative Dragon Kings (nDK) respectively \cite{pisarenko2012robust}.

\emph{The main result of this paper is that the largest values of RV are in fact nDK}. We find that daily returns are the closest to the BS behavior. However, with the increase of $n$ we observe the development of "potential" DK with statistically significant deviations upward from the straight line. This trend terminates with the data points returning to the straight line and then abruptly plunging into nDK territory. 

To gain further insight into this phenomenon, we start in Sec. \ref{TSRV} with the time series of RV from 1970 to 2021, including expanded views of the aforementioned periods of market upheavals. In Sec. \ref{GBDF} we give analytical expressions of the two distribution functions used to fit the entire RV distribution: modified Generalized Beta (mGB), which is discussed in great detail in a companion paper \cite{liu2023rethinking}, and Generalized Beta Prime (GB2), which is essentially a limiting case of mGB and is chosen because it has power-law tails. mGB is chosen because it exhibits long stretch of power-law dependence before dropping off and terminating at a finite value of the variable, thus mimicking the nDK behavior of RV \cite{liu2023rethinking}. Additionally, both mGB and GB2 emerge as steady-state distributions of a stochastic differential equation for stochastic volatility \cite{liu2023rethinking}. In Sec. \ref{RVfit} we describe fits of RV with mGB and GB2 and give a detailed description of the tails, specifically in regards to possible DK/nDK. Towards this end we also use a linear fit (LF) of the tails. For all three fits, we provide confidence intervals \cite{janczura2012black} and, more importantly, the results of a U-test \cite{pisarenko2012robust}, which evaluates a $p$-value for the null hypothesis that a data point comes from a fitting distribution \cite{pisarenko2012robust}. Sec. \ref{Discussion} is a discussion of results obtained in Sec. \ref{RVfit}.

\section{Time Series of Realized Volatility \label{TSRV}}

\begin{figure}[ht]
\centering
\begin{tabular}{c}
\includegraphics[width = .77 \textwidth]{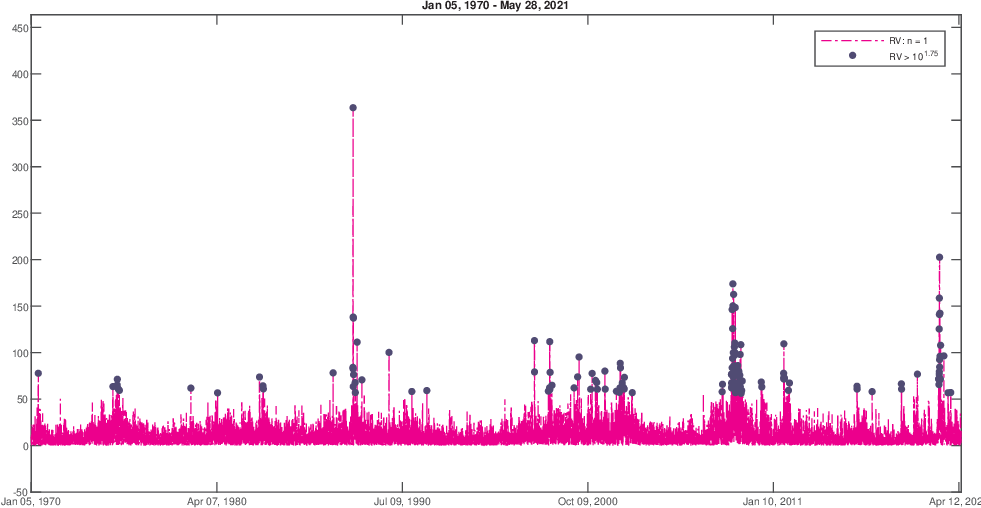}\\
\\
\includegraphics[width = .77 \textwidth]{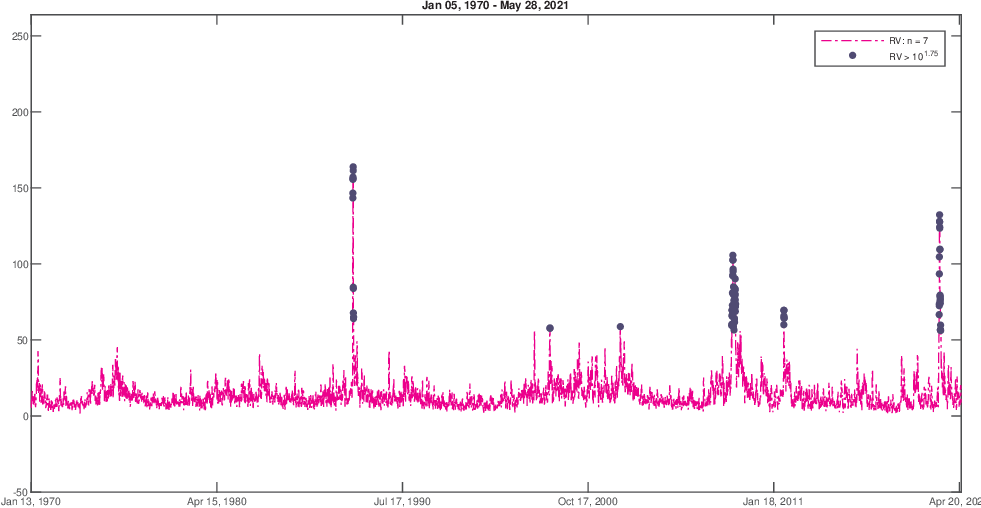}\\
\\
\includegraphics[width = .77 \textwidth]{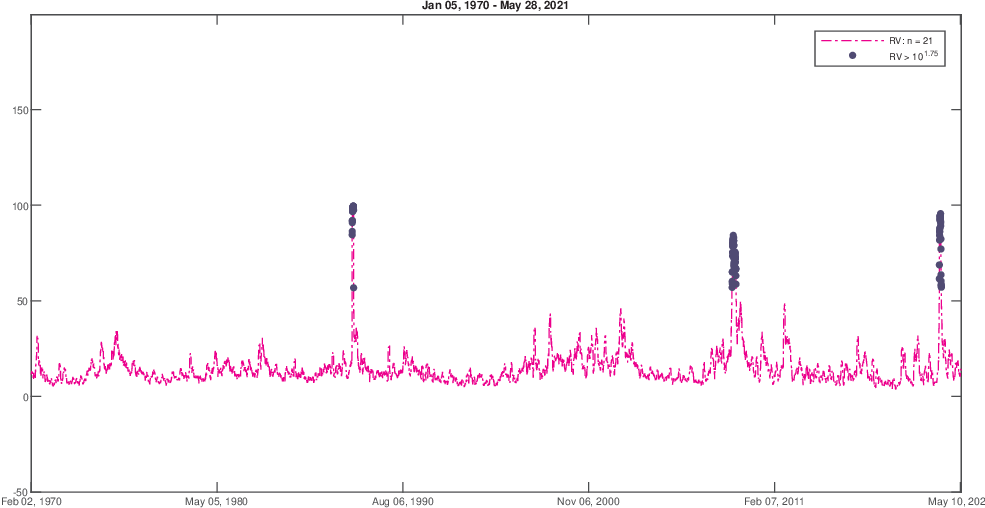}
\end{tabular}
\caption{Time series of RV, with $RV>17$ shown. From top to bottom $n=1,7,21$. Black dots indicate $RV>10^{1.75} \approx 56$.}
\label{TimeSeries}
\end{figure}

Fig. \ref{TimeSeries} shows the time series of RV for $n=1$, based on daily returns, $n=7$, and $n=21$, where $n$ is the number of days over which daily RV is averaged in (\ref{RV2}) and (\ref{RV2av}). Only values with $RV>17$ are shown and black dots mark values $RV>10^{1.75} \approx 56$. It is clear that the time series progression with the increase of $n$ is towards a more pronounced amplification of singularly important events, such as periods corresponding to Black Monday, Financial Crisis and COVID pandemic -- even though the maximum values of RV understandably decrease in the same progression as averaging is taken over a larger number of days $n$. While such progression naturally leads to a question of whether these events might belong to the DK category, we shall see in what follows that they are actually nDK.

\begin{figure}[ht]
\centering
\begin{tabular}{c}
\includegraphics[width = 1.0 \textwidth]{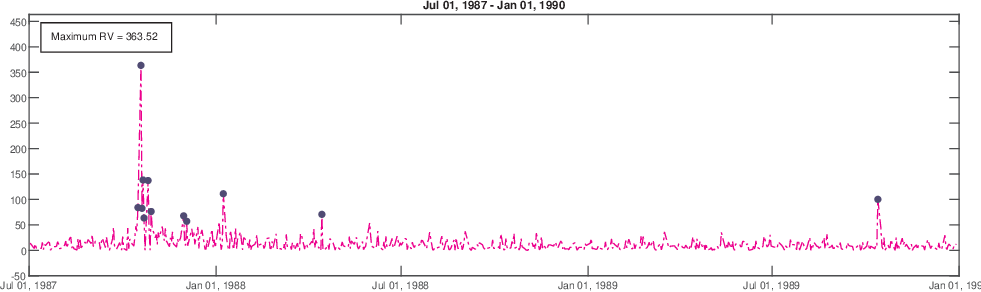}\\
\includegraphics[width = 1.0 \textwidth]{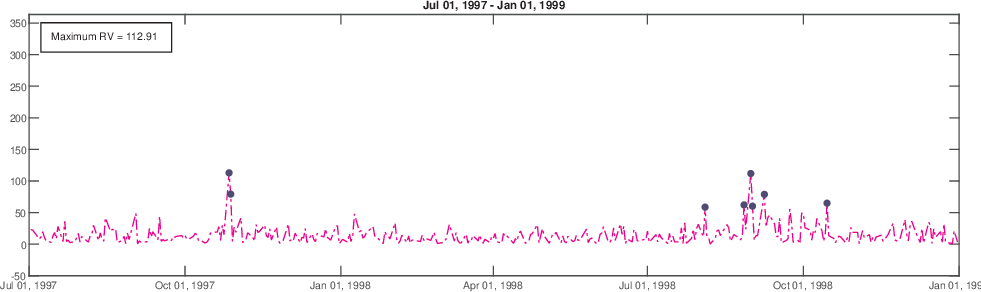}\\
\includegraphics[width = 1.0 \textwidth]{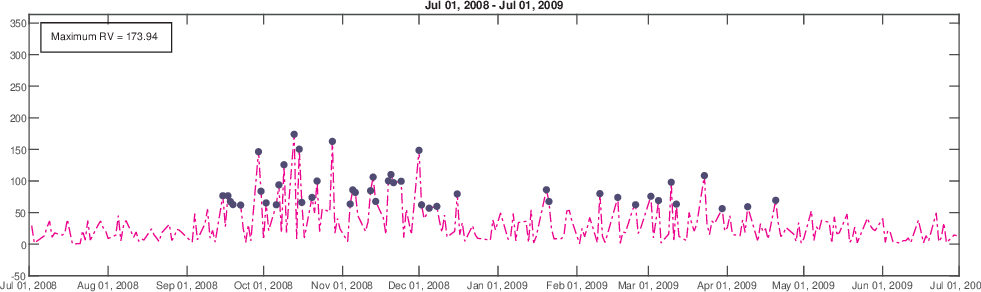}\\
\includegraphics[width = 1.0 \textwidth]{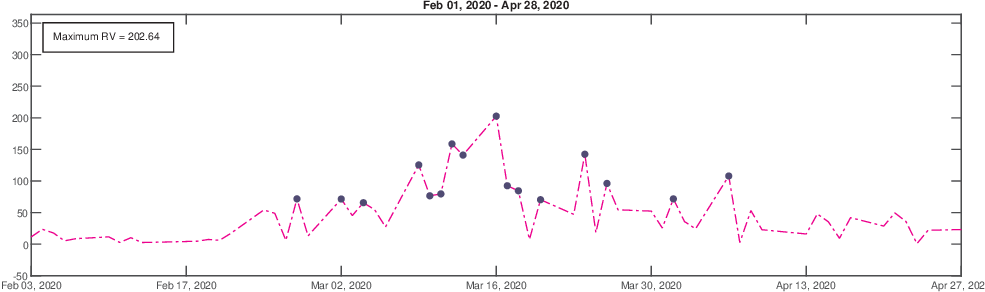}
\end{tabular}
\caption{Snapshots of $n=1$ time series in Fig. \ref{TimeSeries} around Black Monday, Tech Bubble, Financial Crisis and Covid Pandemic.}
\label{TimeSeries_1}
\end{figure}

\begin{figure}[ht]
\centering
\begin{tabular}{c}
\includegraphics[width = 1.0 \textwidth]{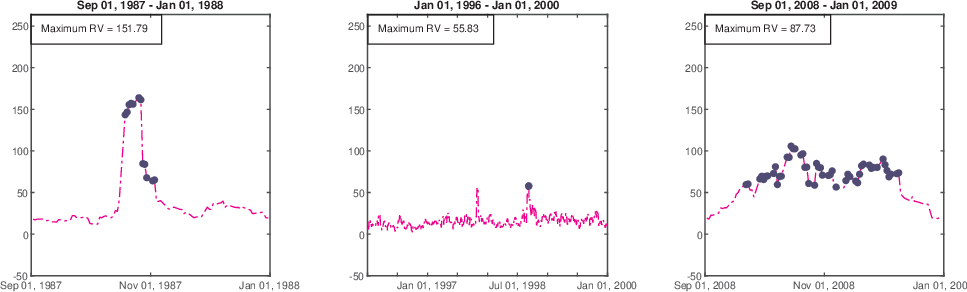}\\
\includegraphics[width = 1.0 \textwidth]{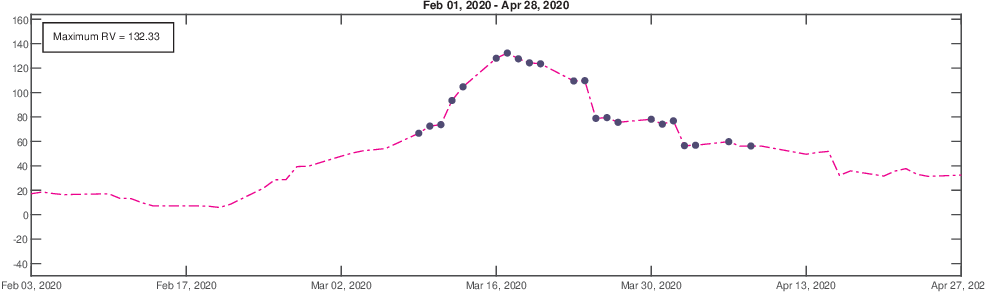}\\
\includegraphics[width = 1.0 \textwidth]{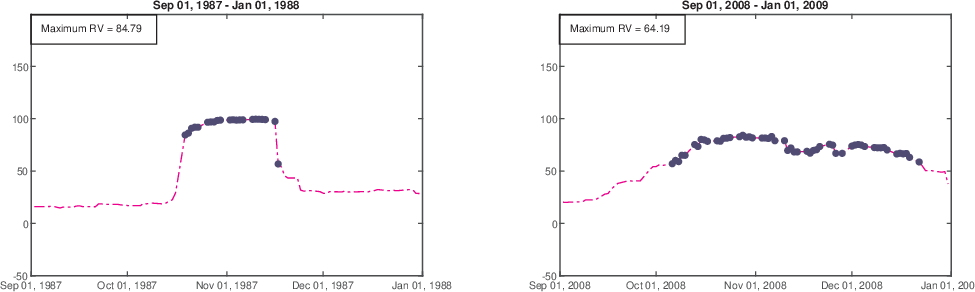}\\
\includegraphics[width = 1.0 \textwidth]{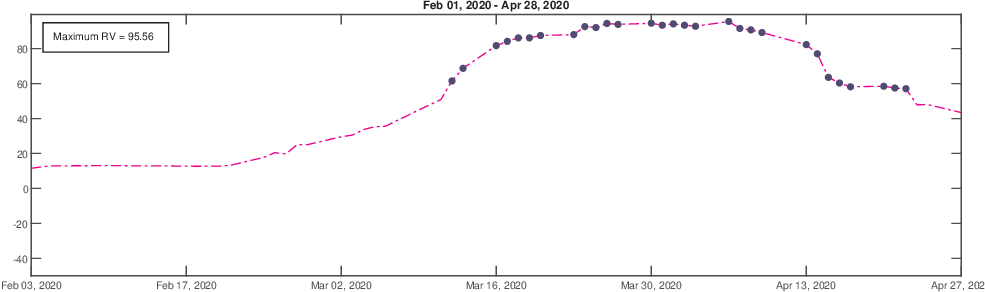}
\end{tabular}
\caption{Snapshots of $n=7$ (top two) and $n=21$ (bottom two) time series in Fig. \ref{TimeSeries} for the same periods as in Fig. \ref{TimeSeries_1}.}
\label{TimeSeries_7_21}
\end{figure}

Figs. \ref{TimeSeries_1} and  \ref{TimeSeries_7_21} give snapshots of the time series in Fig. \ref{TimeSeries} around the largest volatility events: Fig. \ref{TimeSeries_1} based on daily returns, $n=1$, and Fig. \ref{TimeSeries_7_21} for $n=7$ and $n=21$ respectively. Based on Figs. \ref{TimeSeries} -- \ref{TimeSeries_7_21}, Black Monday was clearly the most singular volatility event, while Financial Crisis and COVID Pandemic were distinguished by more prolonged periods of sustained extraordinarily large RV.

\clearpage
\newpage

\clearpage
\newpage

\clearpage
\newpage

\section{Generalized Beta Distribution Function \label{GBDF}}

A companion paper \cite{liu2023rethinking} discusses in great detail the modified Generalized Beta distribution function (mGB) used here to fit the distributions of RV. A generalization of the traditional \cite{mcdonald1995generalization} GB can be written, in a slightly modified form relative to that of \cite{liu2023rethinking}, as follows:
\begin{equation}
f_{GB}(x;\alpha,\beta _1,\beta _2, p,q)= \frac{\alpha \left(1+\left(\frac{\beta _2}{\beta _1}\right)^{\alpha }\right)^p \left(\frac{x}{\beta _2}\right)^{\alpha  p-1} \left(1+\left(\frac{x}{\beta _2}\right)^{\alpha}\right)^{-p-q} \left(1-\left(\frac{x}{\beta _1}\right)^{\alpha }\right)^{q-1}}{\beta _2 B(p,q)},
\label{GBPDF2}
\end{equation}
where $\beta _1$ and $\beta _2$ are scale parameters and $\alpha$, $p$ and $q$ are shape parameters, all positive, $B(p,q)$ is the beta function and $x \leq \beta _1$. Although it has a concise and transparent form, it does not come out as a solution of a stochastic differential equation (SDE) \cite{hertzler2003classical}, which is desirable for the purpose of modeling behavior of quantities, such as stochastic volatility, important for understanding of RV \cite{dashti2021combined}.
 
The probability density function (PDF) of mGB, which comes out as a solution of an SDE (with minor caveats explained in \cite{liu2023rethinking}) and which is used here to model the RV distribution, can be written as 
\begin{equation}
f_{mGB}(x;\alpha,\beta _1,\beta _2, p,q)=\frac{\alpha(p+q)\left(1+\left(\frac{\beta _2}{\beta _1}\right)^\alpha\right)^{p+1}  \left(\frac{x}{\beta _2}\right)^{\alpha p-1}  \left(1+\left(\frac{x}{\beta _2}\right)^\alpha\right)^{-p-q-1} \left(1-\left(\frac{x}{\beta _1}\right)^\alpha\right)^{q-1}}{\beta _2 B(p,q)\left(q+\left(\frac{\beta _2}{\beta _1}\right)^\alpha (p+q) \right) },
\label{mGBPDF}
\end{equation}
The cumulative distribution function (CDF) and complimentary CDF (CCDF) of mGB are given respectively by 
\begin{equation}
\resizebox{1.0\hsize}{!}{$
F_{mGB}(x;\alpha,\beta _1,\beta _2, p,q)=I\left({\frac{\left(\frac{x}{\beta _1}\right)^\alpha+\left(\frac{x}{\beta _2}\right)^\alpha}{1+\left(\frac{x}{\beta _2}\right)^\alpha}};p,q\right) + \frac{1}{B(p,q) \left(q+\left(\frac{\beta _2}{\beta _1}\right)^{\alpha } (p+q)\right)} \left(\frac{1-\left(\frac{x}{\beta _1}\right)^\alpha}{1+\left(\frac{x}{\beta _2}\right)^\alpha}\right)^q\left(\frac{\left(1+\left(\frac{\beta _2}{\beta _1}\right)^\alpha\right) \left(\frac{x}{\beta _2}\right)^\alpha}{\left(1+\left(\frac{x}{\beta _2}\right)^\alpha\right)}\right)^p,
$}
\label{mGBCDF}
\end{equation}
and
\begin{equation}
\resizebox{1.0\hsize}{!}{$
1-F_{mGB}(x;\alpha,\beta _1,\beta _2, p,q)=I\left({\frac{1-\left(\frac{x}{\beta _1}\right)^{\alpha }}{1+\left(\frac{x}{\beta _2}\right)^{\alpha }}};q,p\right) -\frac{1}{B(p,q) \left(q+\left(\frac{\beta _2}{\beta _1}\right)^{\alpha } (p+q)\right)} \left(\frac{1-\left(\frac{x}{\beta _1}\right)^\alpha}{1+\left(\frac{x}{\beta _2}\right)^\alpha}\right)^q\left(\frac{\left(1+\left(\frac{\beta _2}{\beta _1}\right)^\alpha\right) \left(\frac{x}{\beta _2}\right)^\alpha}{\left(1+\left(\frac{x}{\beta _2}\right)^\alpha\right)}\right)^p,
$}
\label{mGBCCDF}
\end{equation}
where the first term in (\ref{mGBCDF}) and (\ref{mGBCCDF}) represent, respectively, CDF and CCDF of GB (whose PDF is given by(\ref{GBPDF2})), while $I(y;p,q)=B(y;p,q)/B(p,q)$ and $B(y;p,q)$ are, respectively, the regularized and incomplete beta functions \cite{nist2022digital}.

In what follows, we will be specifically interested in the $\beta_2\ll\beta_1$ circumstance since for $\beta_2 \ll x \ll \beta_1$ GB and mGB exhibit a power-law dependence,
\begin{equation}
f_{mGB} \propto \left(\frac{x}{\beta _2}\right)^{-\alpha (q+1) - 1}, \hspace{.3cm} 1-F_{mGB} \propto \left(\frac{x}{\beta _2}\right)^{-\alpha (q+1)};
\hspace{.5cm} f_{GB} \propto \left(\frac{x}{\beta _2}\right)^{-\alpha q - 1}, \hspace{.3cm} 1-F_{GB} \propto \left(\frac{x}{\beta _2}\right)^{-\alpha q}.  
\label{GB2tail}
\end{equation}
In the limit of $\beta_1 \rightarrow \infty$, mGB and GB become, respectively, mGB2 and GB2 (the latter also known as Generalized Beta Prime) and are given by \cite{liu2023rethinking}
\begin{equation}
f_{mGB2}(x; \alpha, \beta_2, p,q)=\frac{\alpha (p+q) \left(\frac{x}{\beta_2}\right)^{\alpha p -1} \left(1+\left({\frac{x}{\beta_2}}\right)^{\alpha}\right)^{-p-q-1}}{q\beta_2 B(p,q)},
\label{mGB2PDF}
\end{equation}
and
\begin{equation}
f_{GB2}(x;\alpha,\beta _1,\beta _2, p,q)=\frac{\alpha\left(\frac{x}{\beta _2}\right)^{\alpha  p-1} \left(1+\left(\frac{x}{\beta _2}\right)^{\alpha}\right)^{-p-q}}{\beta _2 B(p,q)}
\label{GB2PDF}.
\end{equation}
Unlike mGB and GB, for whom the power-law dependences in (\ref{GB2tail}) eventually terminate at $\beta_1$, mGB2 and GB2 will sustain these power-law dependences indefinitely.

Below, we will use (\ref{mGBCCDF}) to fit CCDF of distributions of RV. As explained in  \cite{liu2023rethinking}, mGB2 and GB2 are equivalent since $q$ and $p$ are independently defined at this level GB family of distributions and $q$ can be shifted by unity in the definition of mGB2/GB2. Consequently, we choose a more familiar CCDF of GB2 
\begin{equation}
1-F_{GB2}(x;\alpha,\beta _1,\beta _2, p,q)=I\left({\frac{1}{1+\left(\frac{x}{\beta _2}\right)^{\alpha }}};q,p\right)
\label{GB2CCDF}
\end{equation}
to fit CCDF of the RV data. Insofar as the main difference between mGB and GB is concerned, it is their behavior near $\beta_1$ in the present context \cite{liu2023rethinking}. Namely,
\begin{equation}
1-F_{GB} \approx \frac{1}{qB(p,q)} \left(\frac{1-\left(\frac{x}{\beta _1}\right)^\alpha}{1+\left(\frac{x}{\beta _2}\right)^\alpha}\right)^q \approx \frac{1}{qB(p,q)} \left(\frac{1-\left(\frac{x}{\beta _1}\right)^\alpha}{1+\left(\frac{\beta _1}{\beta _2}\right)^\alpha}\right)^q,
\label{GBCDFbeta1}
\end{equation}
and
\begin{equation}
1-F_{mGB} \approx \frac{1+\frac{p}{q}}{qB(p,q)} \left(\frac{1-\left(\frac{x}{\beta _1}\right)^\alpha}{1+\left(\frac{x}{\beta _2}\right)^\alpha}\right)^q \left(\frac{\beta_2}{\beta_1} \right)^{\alpha} \approx \frac{1+\frac{p}{q}}{qB(p,q)} \left(\frac{1-\left(\frac{x}{\beta _1}\right)^\alpha}{1+\left(\frac{\beta _1}{\beta _2}\right)^\alpha}\right)^q \left(\frac{\beta_2}{\beta_1} \right)^{\alpha},
\label{mGBCDFbeta1}
\end{equation}
that is $1-F_{mGB}$ drops off to zero ($F_{mGB}$ saturates to unity) faster than $1-F_{GB}$ due to the factor $\left(\frac{\beta _2}{\beta _1}\right)^\alpha$. This feature accounts for a better fit via mGB versus GB, which may be due to the fact that mGB emerges from a physically motivated stochastic model \cite{liu2023rethinking}.

\section{Fitting Distribution of Realized Volatility  \label{RVfit}}

\begin{figure}[!htbp]
\centering
\begin{tabular}{c}
\includegraphics[width = .77 \textwidth]{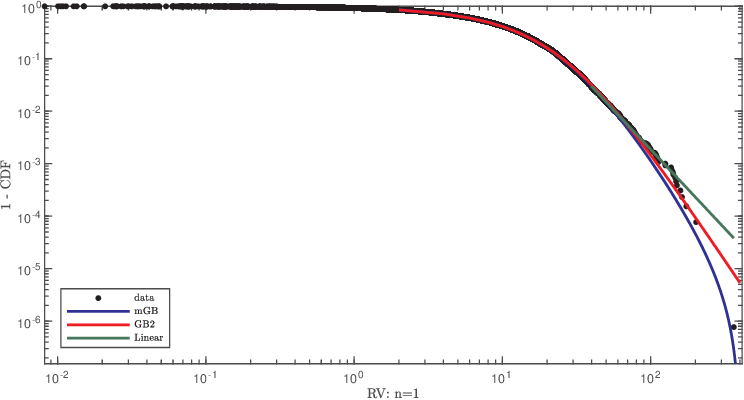}\\
\\
\includegraphics[width = .77 \textwidth]{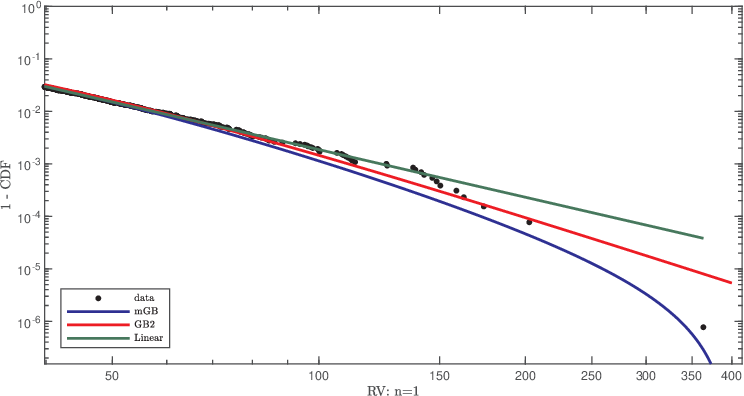}\\
\\
\includegraphics[width = .77 \textwidth]{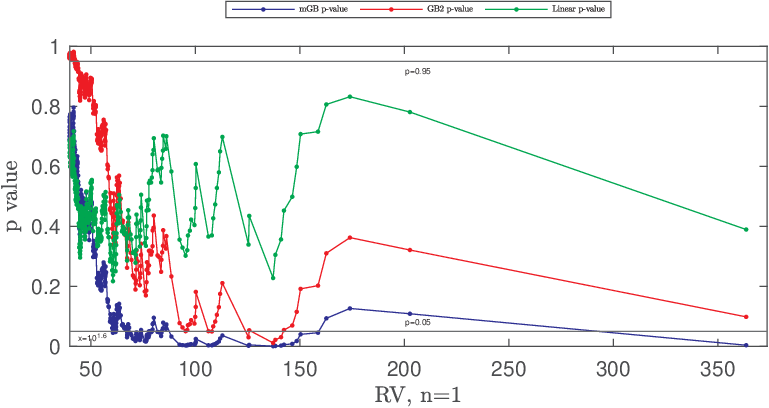}\\
\end{tabular}
\caption{Linear, GB2 and mGB fits (top), tail, $RV>40$ (middle) and p-values (bottom) for $n=1$.} 
\label{RVallfits_1}
\end{figure}

\begin{figure}[p]
\centering
\begin{tabular}{c}
\includegraphics[width = .77 \textwidth]{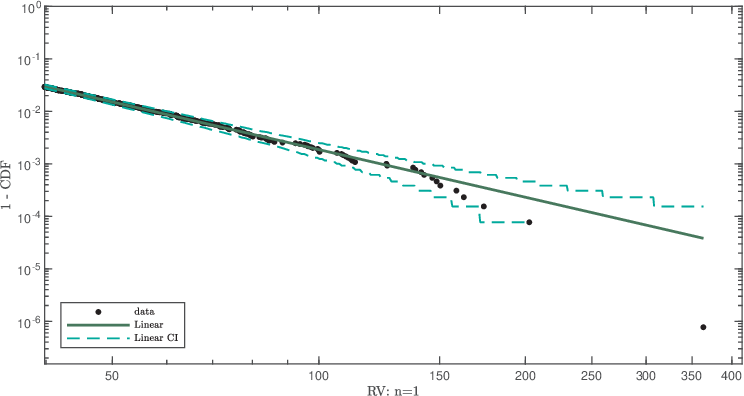}\\
\\
\includegraphics[width = .77 \textwidth]{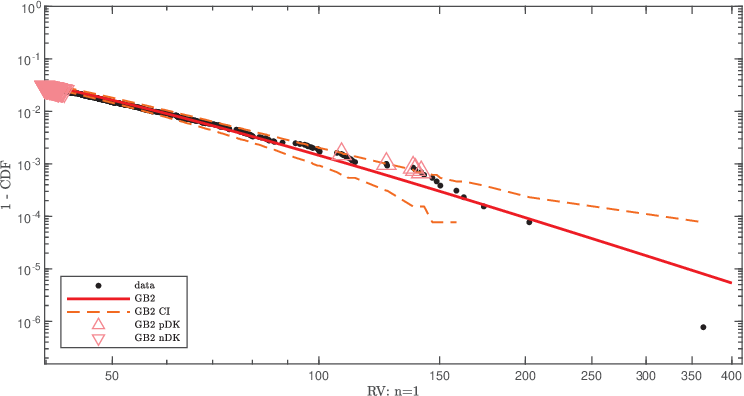}\\
\\
\includegraphics[width = .77 \textwidth]{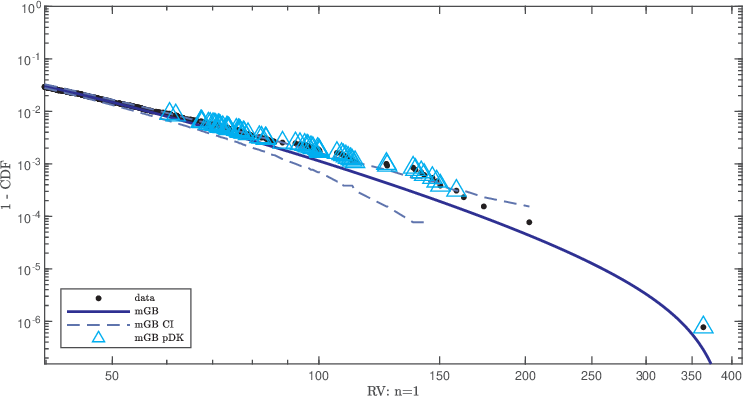}
\end{tabular}
\caption{Same as the middle Fig. \ref{RVallfits_1} with the respective CI, "potential" DK (up triangles) and nDK (down triangles).}
\label{RVallfitstail_1}
\end{figure}

\begin{figure}[!htbp]
\centering
\begin{tabular}{c}
\includegraphics[width = .77 \textwidth]{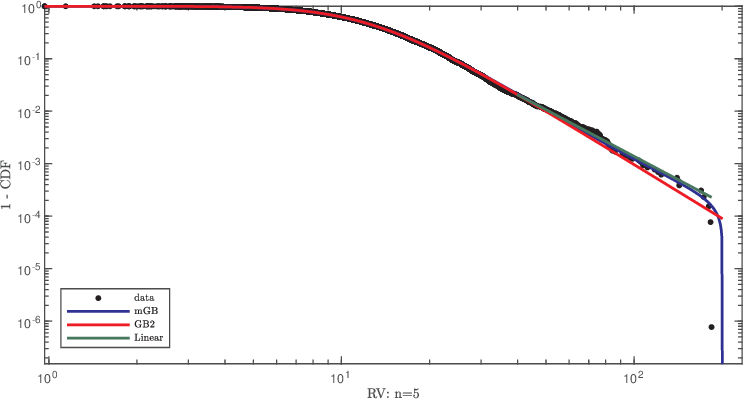}\\
\\
\includegraphics[width =.77 \textwidth]{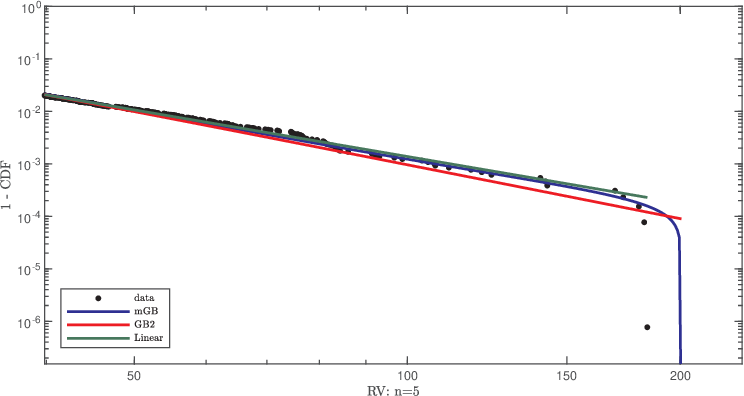}\\
\\
\includegraphics[width = .77 \textwidth]{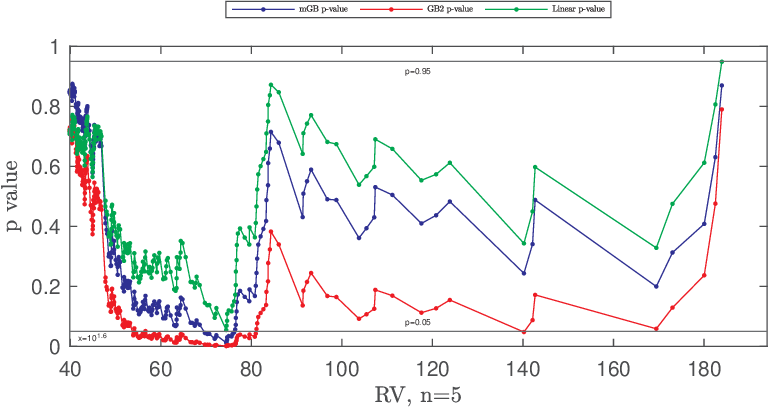}\\
\end{tabular}
\caption{Linear, GB2 and mGB fits (top), tail, $RV>40$ (middle) and p-values (bottom) for $n=5$.}
\label{RVallfits_5}
\end{figure}

\begin{figure}[p]
\centering
\begin{tabular}{c}
\includegraphics[width = .77 \textwidth]{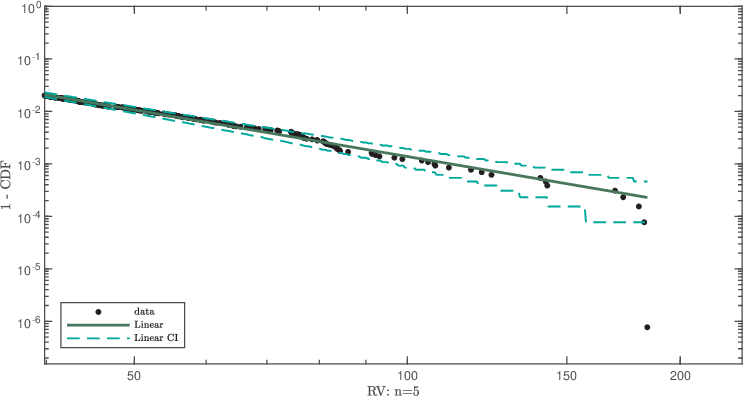}\\
\\
\includegraphics[width = .77 \textwidth]{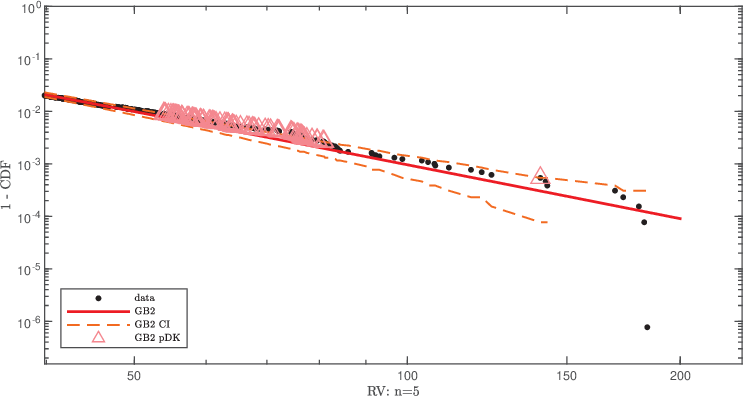}\\
\\
\includegraphics[width = .77 \textwidth]{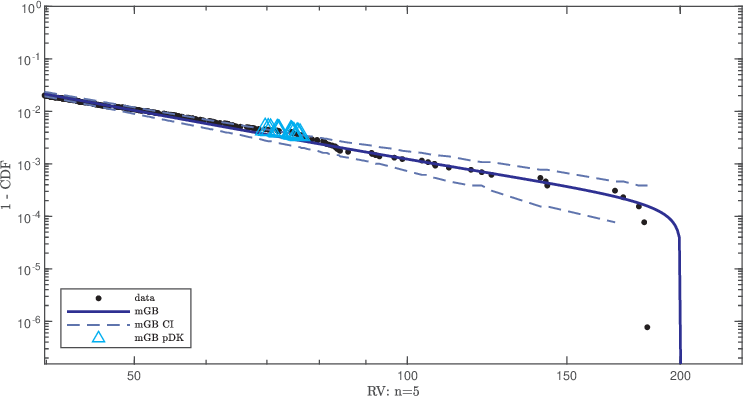}
\end{tabular}
\caption{Same as the middle Fig. \ref{RVallfits_5} with the respective CI, "potential" DK (up triangles) and nDK (down triangles).}
\label{RVallfitstail_5}
\end{figure}

\begin{figure}[!htbp]
\centering
\begin{tabular}{c}
\includegraphics[width = .77 \textwidth]{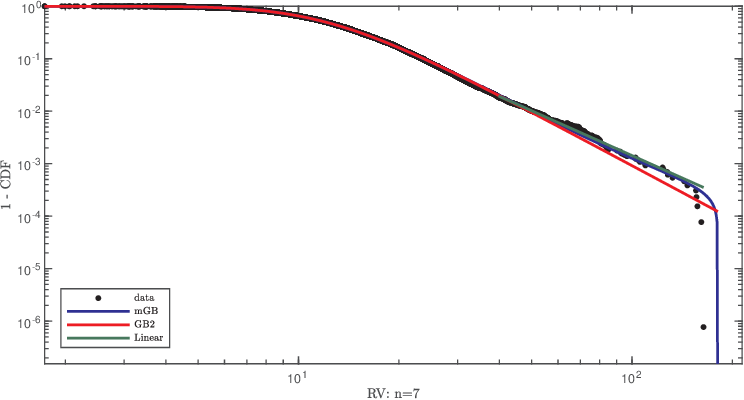}\\
\\
\includegraphics[width =.77 \textwidth]{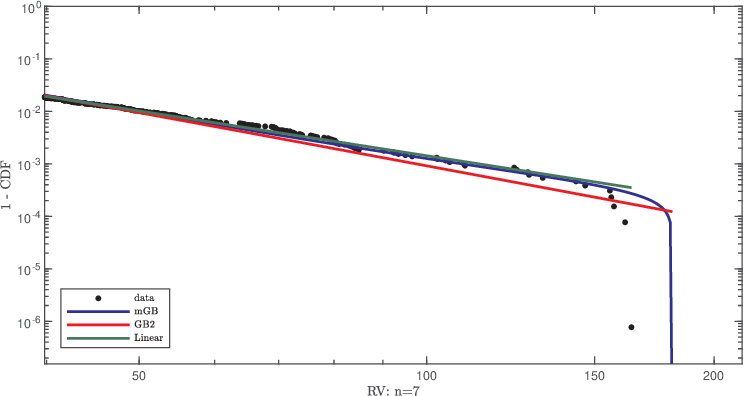}\\
\\
\includegraphics[width = .77 \textwidth]{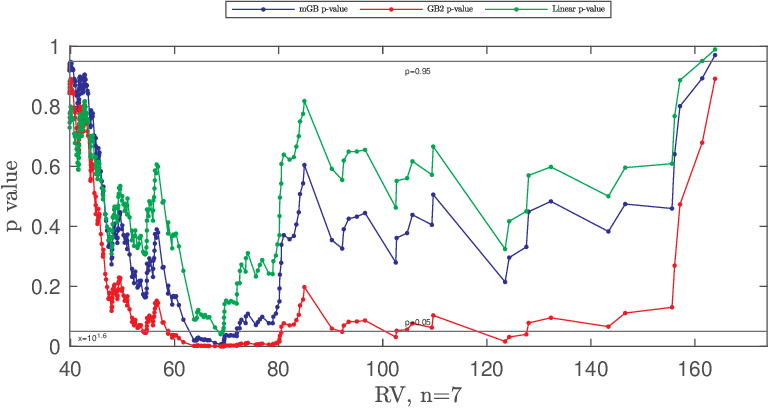}\\
\end{tabular}
\caption{Linear, GB2 and mGB fits (top), tail, $RV>40$ (middle) and p-values (bottom) for $n=7$.}
\label{RVallfits_7}
\end{figure}

\begin{figure}[p]
\centering
\begin{tabular}{c}
\includegraphics[width = .77 \textwidth]{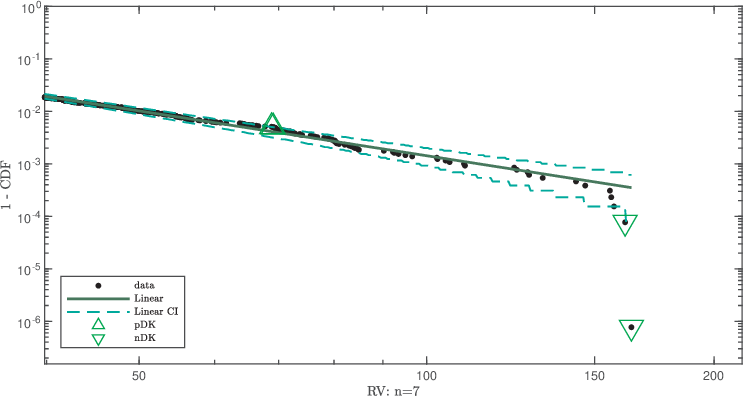}\\
\\
\includegraphics[width = .77 \textwidth]{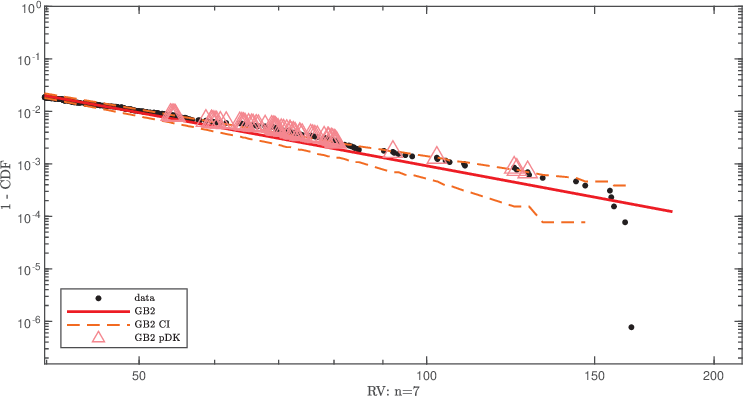}\\
\\
\includegraphics[width = .77 \textwidth]{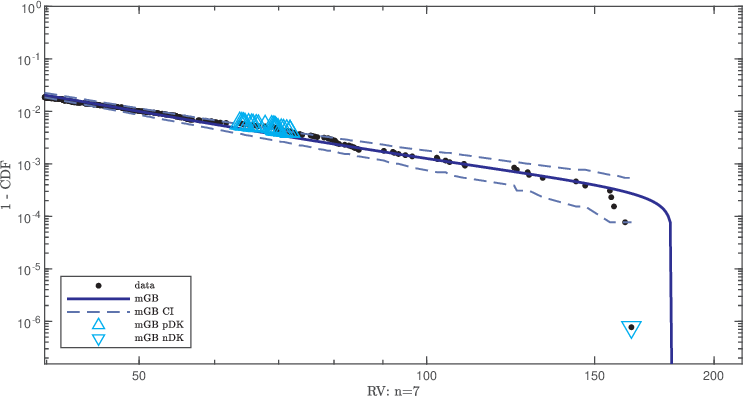}
\end{tabular}
\caption{Same as the middle Fig. \ref{RVallfits_7} with the respective CI, "potential" DK (up triangles) and nDK (down triangles).}
\label{RVallfitstail_7}
\end{figure}

\begin{figure}[!htbp]
\centering
\begin{tabular}{c}
\includegraphics[width = .77 \textwidth]{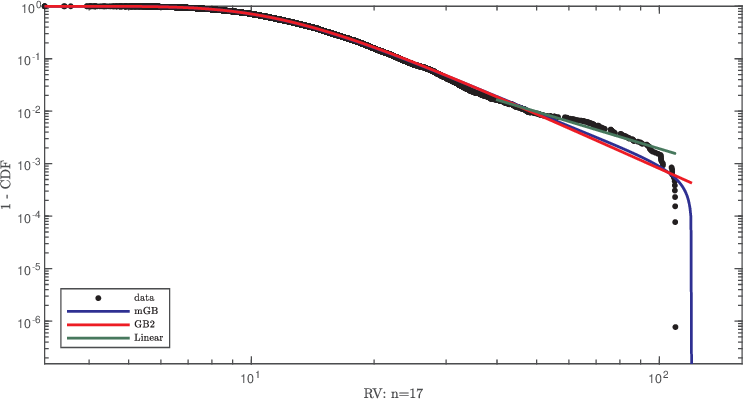}\\
\\
\includegraphics[width =.77 \textwidth]{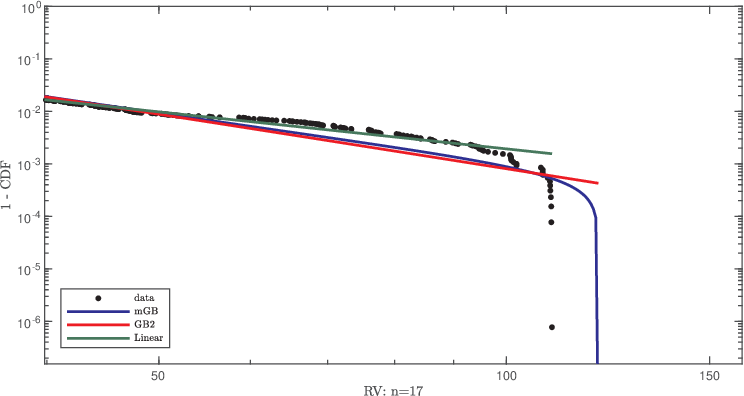}\\
\\
\includegraphics[width = .77 \textwidth]{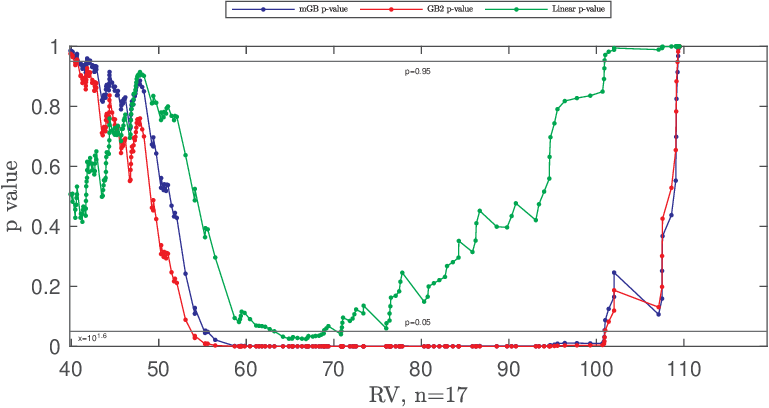}\\
\end{tabular}
\caption{Linear, GB2 and mGB fits (top), tail, $RV>40$ (middle) and p-values (bottom) for $n=17$.}
\label{RVallfits_17}
\end{figure}

\begin{figure}[p]
\centering
\begin{tabular}{c}
\includegraphics[width = .77 \textwidth]{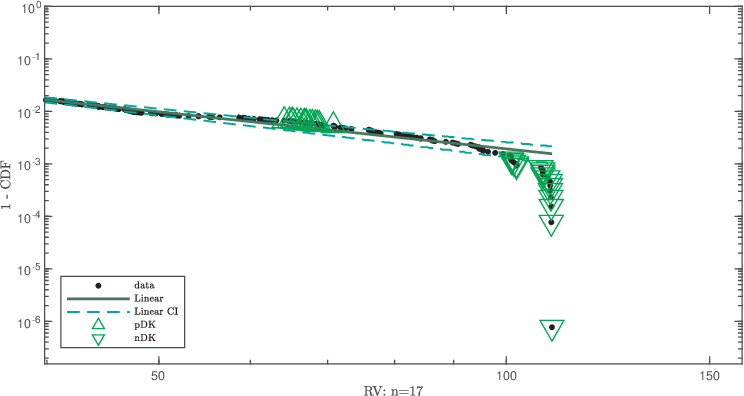}\\
\\
\includegraphics[width = .77 \textwidth]{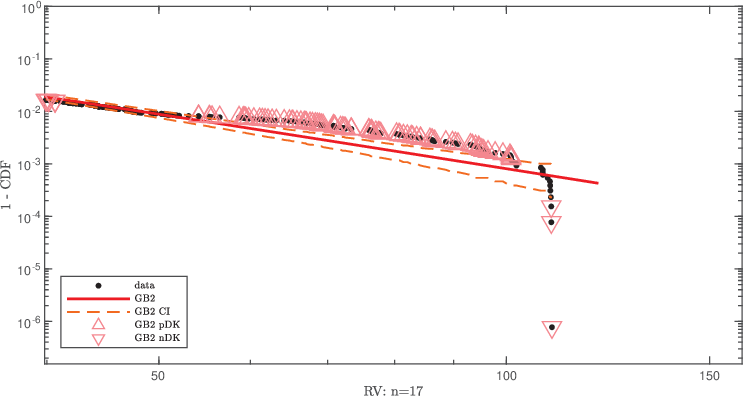}\\
\\
\includegraphics[width = .77 \textwidth]{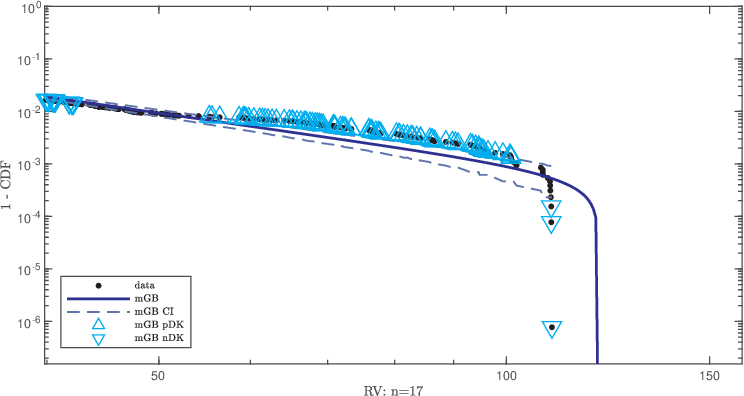}
\end{tabular}
\caption{Same as the middle Fig. \ref{RVallfits_21} with the respective CI, "potential" DK (up triangles) and nDK (down triangles).}
\label{RVallfitstail_17}
\end{figure}

\begin{figure}[!htbp]
\centering
\begin{tabular}{c}
\includegraphics[width = .77 \textwidth]{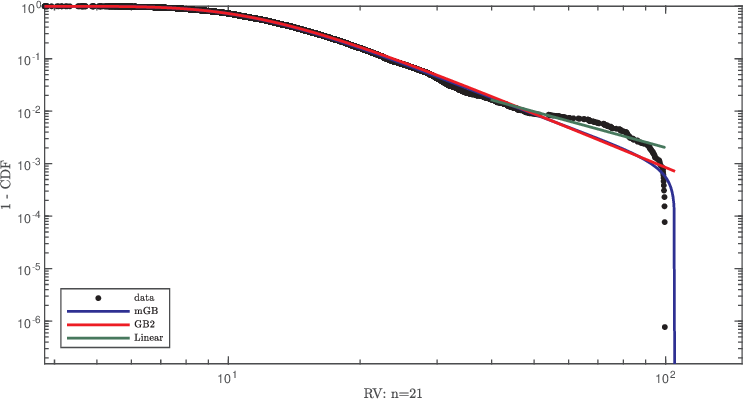}\\
\\
\includegraphics[width =.77 \textwidth]{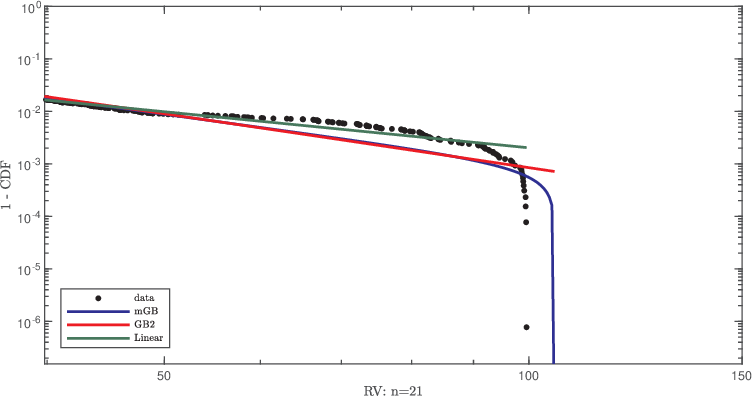}\\
\\
\includegraphics[width = .77 \textwidth]{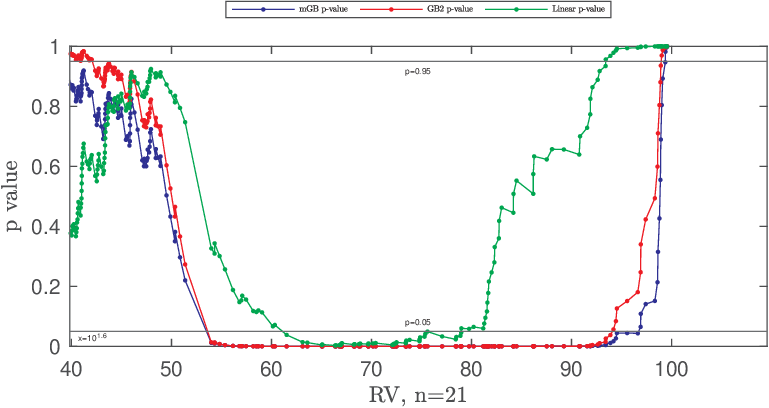}\\
\end{tabular}
\caption{Linear, GB2 and mGB fits (top), tail, $RV>40$ (middle) and p-values (bottom) for $n=21$.}
\label{RVallfits_21}
\end{figure}

\begin{figure}[p]
\centering
\begin{tabular}{c}
\includegraphics[width = .77 \textwidth]{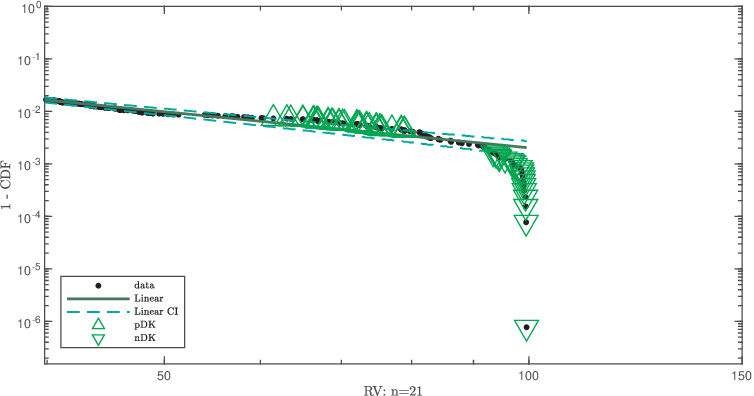}\\
\\
\includegraphics[width = .77 \textwidth]{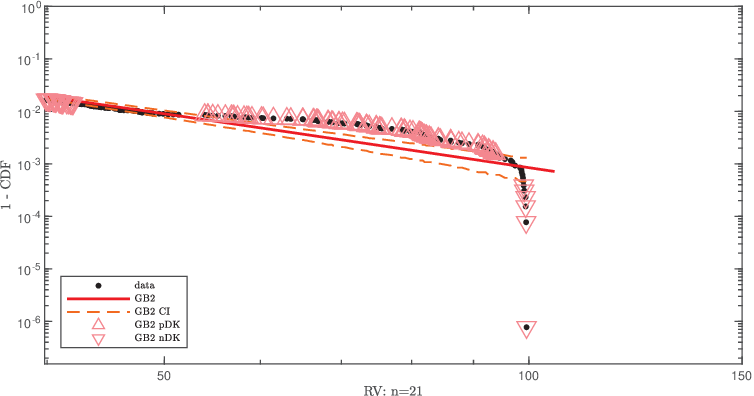}\\
\\
\includegraphics[width = .77 \textwidth]{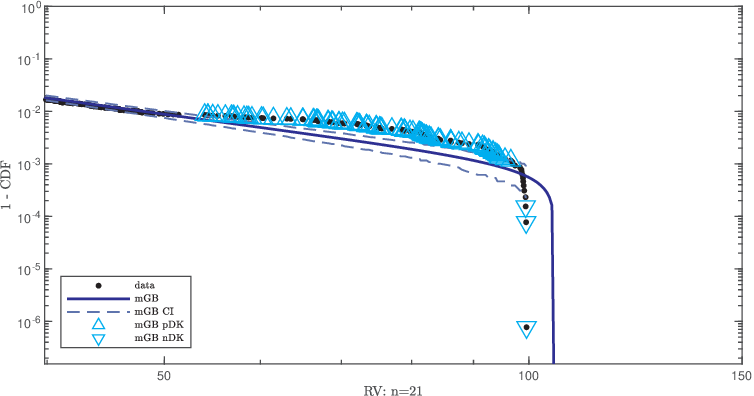}
\end{tabular}
\caption{Same as the middle Fig. \ref{RVallfits_21} with the respective CI, "potential" DK (up triangles) and nDK (down triangles).}
\label{RVallfitstail_21}
\end{figure}

We fit CCDF of the full RV distribution -- for the entire time span discussed in Sec. \ref{TSRV} -- using mGB (\ref{mGBCCDF}) and GB2 (\ref{GB2CCDF}). The fits are shown on the log-log scale in Figs. \ref{RVallfits_1} -- \ref{RVallfitstail_21}, together with the linear fit (LF) of the tails with $RV>40$. LF excludes the end points, as prescribed in \cite{pisarenko2012robust}, that visually may be nDK candidates. (In order to mimic LF we also excluded those points in GB2 fits, which has minimal effect on GB2 fits, including the slope and KS statistic). To make the progression of the fits as a function of $n$ clearer, we included results for $n=5$ and $n=17$, in addition to $n=1,7,21$ that we used in Sec. \ref{TSRV}. Confidence intervals (CI) were evaluated per \cite{janczura2012black}, via inversion of the binomial distribution. $p$-values were evaluated in the framework of the U-test, which is discussed in \cite{pisarenko2012robust} and is based on order statistics:
\begin{equation}
p(x_{k,n})=1-B\left(F(x_{k,n}); k, n-k+1\right),
\label{p-value}
\end{equation}
where $x_{k,n}$ is the $k$'s member of numbers between $1$ and $n$ ordered by increasing  magnitude (RV values in this case), and $F(x_{k,n})$ is the assumed CDF (mGB, GN2 and LF here).

\begin{figure}[p]
\centering
\begin{tabular}{c}
\includegraphics[width = .77 \textwidth]{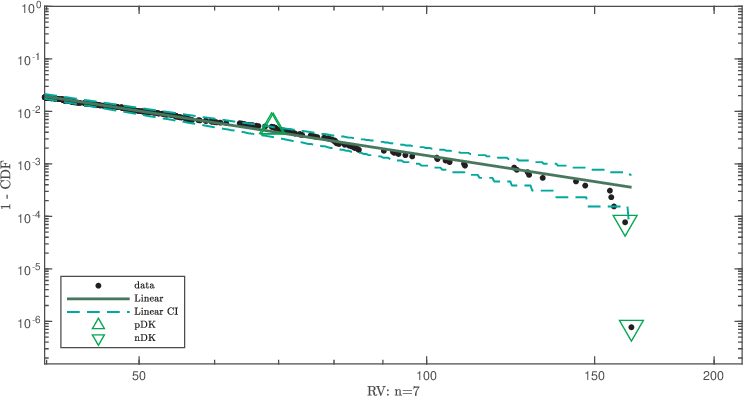}\\
\\
\includegraphics[width = .77 \textwidth]{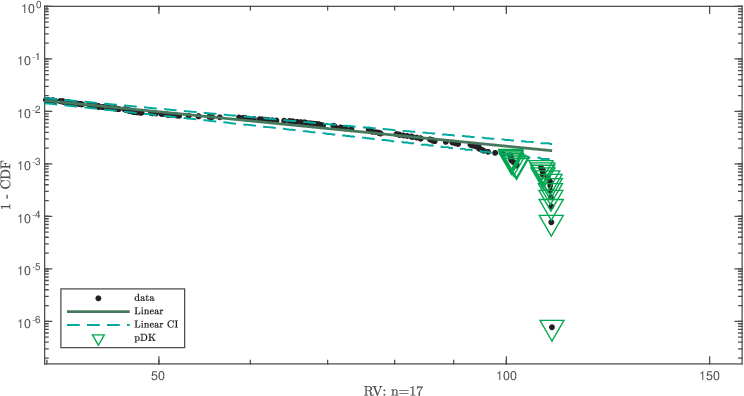}\\
\\
\includegraphics[width = .77 \textwidth]{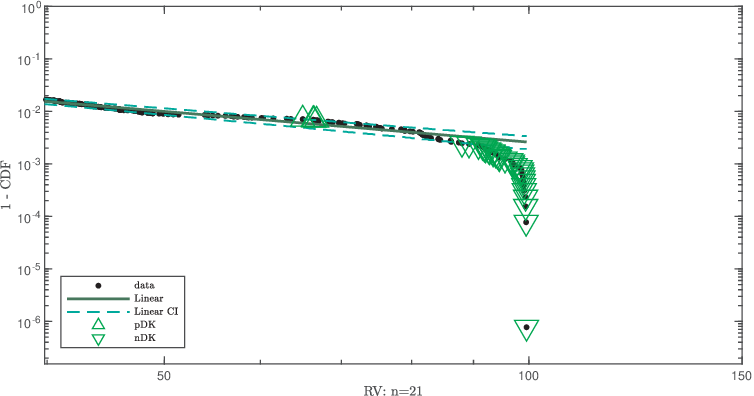}
\end{tabular}
\caption{From top to bottom, LF for $n=7,17,21$ with end values $>0.9\max\{RV\}$ excluded from LF.}
\label{RVLF10pc_7_17_21}
\end{figure}

\begin{figure}[htbp!]
	\centering
		\includegraphics[width = .42 \textwidth]{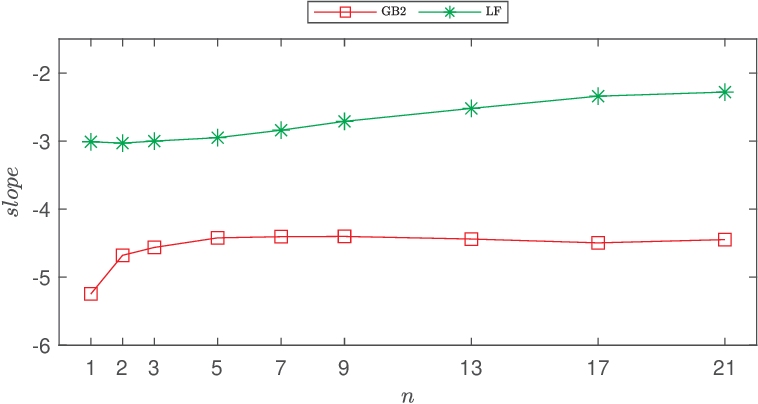} \hspace{.35cm}
		\includegraphics[width = .42 \textwidth]{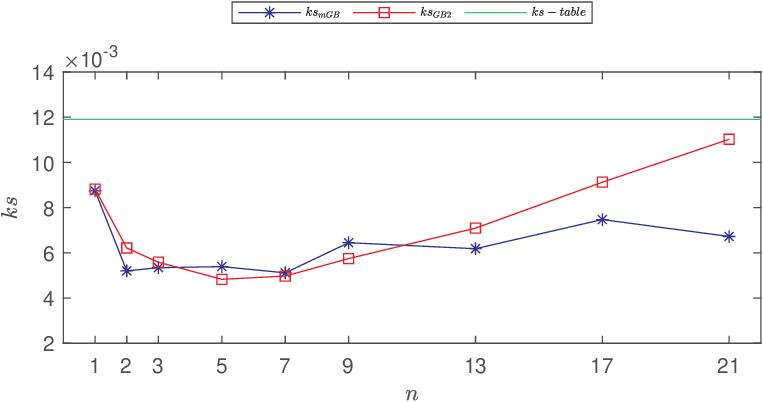}		
	\caption{As a function of $n$: linear slope values of linear (stars) and GB2 (squares) fits -- left; KS statistic values of mGB (stars) and GB2 (squares) fits -- right}
	\label{n_dependence}
\end{figure}

For each $n$, from top to bottom, Figs. \ref{RVallfits_1} -- \ref{RVallfitstail_21} are organized as follows:
\begin{itemize}
\item Full data CDF fit with mGB and GB2 and LF of the tails;
\item Same as above shown for $RV>40$;
\item $p$-values of all three fits for $RV>40$, with $p<0.05$ indicating DK and $p>0.95$ nDK;
\item LF with its CI;
\item GB2 fit with its CI;
\item mGB fit with its CI.
\end{itemize}
In the CI plots, upward pointing triangles indicate $p$-values consistent with DK, while downward pointing triangles indicate $p$-values consistent with nDK \cite{pisarenko2012robust}. Fig. \ref{RVLF10pc_7_17_21} shows LF for $n=7,17,21$, where the last 10\% of the range of values were excluded, that is values greater than $0.9\max\{RV\}$, as opposed to excluding points visually as in Figs. \ref{RVallfits_1} -- \ref{RVallfitstail_21}. 
Fig. \ref{n_dependence} shows LF and GB2 slopes and Kolmogorv-Smirnov (KS) statistic for GB2 and mGB as a function of $n$; the horizontal line with the table value of KS statistic for our sample size \cite{massey1985kolmogorov} is shown for guidance only since mGB and GB2 here are distributions with estimated parameters. 

\clearpage
\newpage

\clearpage
\newpage


\clearpage
\newpage

\clearpage
\newpage


\clearpage
\newpage

\clearpage
\newpage


\clearpage
\newpage

\clearpage
\newpage


\clearpage
\newpage

\clearpage
\newpage

\clearpage
\newpage

\section{Discussion\label{Discussion}}
While the standard search for Dragon Kings involves performing a linear fit of the tails of the distribution \cite{pisarenko2012robust,janczura2012black}, here we tried to broaden our analysis by also fitting the entire distribution using mGB (\ref{mGBCCDF}) and GB2 (\ref{GB2CCDF}) -- the two members of the Generalized Beta family of distributions \cite{liu2023rethinking}, \cite{mcdonald1995generalization}. As explained in the paragraph that follows (\ref{mGBCCDF}), the central feature of mGB is that, after exhibiting a long power-law dependence, it eventually terminates at a finite value of the variable. GB2, on the other hand, has a power-law tail that extends mGB's power-law dependence to infinity.

The key to understanding the results of fits in Sec. \ref{RVfit} is the analysis of the structure of RV used by the markets -- a square root of realized variance (\ref{RV2}). At its core is the average of the consecutive daily realized variances (\ref{RV2av}). Distribution of daily realized variance can be modeled using a duo of stochastic differential equations -- for stock returns and stochastic volatility -- which produces distributions of daily variance such as mGB \cite{liu2023rethinking} and GB2 \cite{dashti2021combined}. Via a simple change of variable, daily RV would then follow the same distributions but with renormalized parameters.

Even assuming the knowledge of the distribution of daily realized variance, finding the distribution of the averages constitutes a daunting task. To begin with, using convolution to evaluate the distribution of a sum of just two such complex distributions as mGB and GB2 is already not amenable to analytical evaluation. To complicate things further, the consecutive daily RV cannot be treated as independent identically distributed variable (i.i.d.) due to the correlations that persists up to roughly 5 - 7 days \cite{dashti2021realized}.

With the above in mind, we first address Figs. \ref{RVallfits_1} --  \ref{RVallfitstail_21}. According to Figs. \ref{RVallfits_1} and \ref{RVallfitstail_1}, daily RV appears to be the closest of being commensurate with the Black Swan behavior as both LF and GB2 approximate the tail of the distribution better than mGB and LF does not point to existence of either DK, $p<0.05$,  or nDK, $p>0.95$. The $n=1$ behavior undergoes a dramatic change with the increase of $n$, as seen in Figs. \ref{RVallfits_5} --  \ref{RVallfitstail_21}, where we observe that, first, the "potential" DK, $p<0.05$, develop at the earlier portions of the tails, only to terminate in nDK at the tail ends. 

Generally speaking, the existence of the large number of "potential" DK in the tail of the distribution indicates that the distribution is not describing the tail adequately. This becomes pronounced for large $n$ for all three fits -- LF, GB2 and mGB -- although less so for LF, which also does not exhibit "potential" DK for small $n$. However, if we adopt a different procedure for LF, whereby instead of visually excluding nDK candidates at the tail end we exclude the values whose RV is greater than $0.9$ of the maximum RV, we observe in Fig. \ref{RVLF10pc_7_17_21} that it has little effect on LF for small $n$ but all but eliminates "potential" DK.

For large $n$ we also observe that mGB approximates the tail end better than GB2 -- consistent with smaller KS values in Fig. \ref{n_dependence} and smaller number of nDK. However, neither approximates the preceding portion of the tail well as indicated by the "potential" DK. This has to do with the fact that neither of the distributions appear as a solution of a first-principle model describing average RV. Finally, in the first plot in Fig. \ref{n_dependence}, we observe that after roughly 5 - 7 days the slope of the GB2 tail saturates, consistent with the correlation range of daily RV \cite{dashti2021realized}. The slope of LF, on the other hand, increases  with $n$. However neither is consistence with a naive assumption of the distribution having the same slope as that of the daily RV. 

In conclusion, we showed that for daily returns distribution of realized volatility likely has a power-law tail, consistent with the Black Swan behavior. Multi-day realized volatility develops strong negative Dragon King signature as the number of days involved in averaging of daily realized variances increases. The breadth and strength of the S\&P index analyzed here may be a contributing factor in suppressing the runaway power-law behavior. A natural extension of this work will be analysis of gains and losses of stock returns as well as of other large data sets that call into question possible power-law tails, such as incomes.

\bibliography{mybib}

\end{document}